\documentclass[11pt]{article}
\usepackage[parfill]{parskip}    
\usepackage{orcidlink}
\usepackage{graphicx}
\usepackage{amssymb}
\usepackage{epstopdf}
\usepackage[T1]{fontenc}
\usepackage{graphicx}
\usepackage{hyperref}
\usepackage{authblk}
\hypersetup{colorlinks=true, allcolors=blue}
\newcommand{\keywords}[1]{\par\addvspace\baselineskip\noindent\textbf{Keywords:} #1}
\title{A Computational Approach to Sustainable Policies Evaluation of the Italian Wheat Production System}
\author[1]{Gianfranco Giuloni \orcidlink{0000-0003-0946-1738}}
\author[2]{Edmondo Di Giuseppe \orcidlink{0000-0002-6443-9106}}
\author[2]{Arianna Di Paola\orcidlink{0000-0001-9050-4787}}
\affil[1]{ Dipartimento di Studi Socio-Economici, Gestionali e Statistici (DiSEGS), Viale Pindaro, 42, 65127 Pescara, Italy\\%
}
\affil[2]{Institute of BioEconomy, National Research Council of Italy (IBE-CNR), Via dei Taurini, 19, 00185 Roma, Italy}
\date{}   
\begin{document}
\maketitle              
\section*{Abstract}

	This work outlines the modeling steps for developing a tool aimed at supporting policymakers in
	guiding policies toward more sustainable wheat production. In the agricultural sector,
	policies affect a highly diverse set of farms, which differ across several dimensions such as size,
	land composition, local climate, and irrigation availability.
	To address this significant heterogeneity, we construct an Agent-Based Model (ABM).
	The model is initialized using a representative survey of Italian farms, which captures their heterogeneity.
	The ABM is then scaled to include a number of farms
	comparable to those operating nationwide.
	To capture broader dynamics, the ABM is integrated with two additional components:
	a global model of international wheat markets and a tool for assessing the environmental impacts
	of wheat production. This integrated framework enables us to account for the feedback loop between
	global prices and local production while evaluating the environmental implications of policy measures.

\keywords{agent-based model\and farm crop management\and yield-gap\and global trade network\and Life Cycle Assessment (LCA)}
\section{Introduction}

Addressing the environmental impact of food production is a key focus in current sustainability research \cite{poore_reducing_2018}. To understand the scale of this issue, food production accounts for over a quarter (26\%) of global greenhouse gas emissions, uses half of the world’s habitable land for agriculture, consumes 70\% of global freshwater withdrawals, and is responsible for 78\% of ocean and freshwater eutrophication. As highlighted by \cite{hannah_ritchie_environmental_2020}, ``Ensuring everyone in the world has access to a nutritious diet in a sustainable way is one of the greatest challenges we face.''

Among staple crops, global wheat production increased by 42\% between 1993 and 2019, despite the harvested area remaining nearly constant. This reflects a significant intensification of agricultural practices, which is associated with well-documented negative environmental impacts. Countries with high-input farming systems, such as Italy, must urgently adopt sustainable green solutions to maintain current wheat production levels while preserving ecosystems \cite{fischer_issues_2018}.

On January 1, 2023, the European Union implemented the Common Agricultural Policy (CAP) for 2023–2027 \cite{greenarch}. This policy is structured around ten specific objectives, aligned with the EU’s broader social, environmental, and economic sustainability goals for agriculture and rural areas. Subsequently, on December 2, 2022, the European Commission approved Italy’s National Strategic Plan (PSP) for 2023–2027 \cite{itstratplan}.

The PSP incorporates a range of environmental measures, including incentives and penalties, mandatory and voluntary actions, and single or coupled schemes. Within this framework, farmers can choose among different combinations of measures.

Moreover, these policies affect an agricultural sector composed of a highly diverse set of farms, which differ in several dimensions such as size and, most importantly, their technical production conditions (land composition, local weather, irrigation availability, and so on).

Handling this variety of policy options across heterogeneous agents is challenging for traditional modeling approaches. A multi-agent simulation appears to be a suitable way to jointly assess the economic and environmental sustainability of farms at the national level.

Numerous studies have already applied agent-based models to evaluate environmental impacts and agricultural policies (see \cite{kremmydas_review_2018} for a review). This approach is particularly valuable for capturing the varied effects of policies across heterogeneous farms (\cite{khan_economic_2020-2,siad_durum_2017-3}).

This study introduces a framework for an agent-based model designed to evaluate the economic and environmental sustainability of wheat production in Italy. Specifically, it examines how farmers’ behaviors adapt to the implementation of green policies by national authorities.

However, one of the most important variables affecting farmer behavior is the wheat price, typically determined on international markets.
The model of the Italian wheat production system is therefore linked with an existing computational tool that simulates international wheat markets. In this way, the two-way interaction between farmers’ choices and international prices is endogenously represented.

The model is also connected with a Life Cycle Assessment Model (LCA) module to evaluate the environmental impact of wheat production.
Altogether, these routines form the integrated modeling framework for the Italian wheat production system, designed to support policymakers.
The paper is organized as follows: Section 2 presents the Agent-Based framework, Section 3 describes the global economic model determining market prices,
and Section 4 focuses on the environmental impact assessment. The final section discusses technical aspects and outlines future research directions.

\section{Modeling the Italian wheat production system}

In this section, we describe the architecture of the integrated modeling framework for the Italian wheat production system. The framework combines three main components: an Agent-Based Model (ABM), a Global Economic Model (GEM), and a Life Cycle Assessment (LCA) module. These elements are dynamically linked to represent the feedback between local farm-level decisions, international wheat markets, and environmental outcomes.

The ABM simulates a synthetic population of farms equal in size to the actual number of durum wheat producers in Italy. Farm heterogeneity is captured using microdata from the RICA/FADN (\url{https://rica.crea.gov.it}) database, which describes differences in farm size, resource use, and production practices. These observations are used to statistically characterize clusters of farms, which are then expanded to populate the ABM with representative agents.

At the start of each simulation cycle, the ABM receives as inputs both international wheat prices—computed by the GEM based on global supply and demand—and the set of policy measures under evaluation. Farmers, represented as agents, react to these external drivers by making production decisions, including yield target, inputs use, and participation in voluntary or mandatory green schemes.

Once each seasonal cycle concludes, the environmental impacts of farmers’ decisions are assessed through the LCA module. The LCA is applied to every farm in the simulation, producing both farm-level and aggregated national indicators of impacts such as greenhouse gas emissions, ecosystem quality, and resource depletion. Simultaneously, the ABM aggregates farm yields to estimate total national wheat production. This aggregated output feeds back into the GEM, which recalculates global market prices based on the updated balance of supply and demand. These new prices are then passed to the ABM as inputs for the following simulation period, closing the feedback loop.

This integrated system enables the evaluation of how Italian wheat farmers adapt to policy changes in a context where international prices and environmental impacts are dynamically connected. By iterating the feedback cycle across multiple simulated seasons, the framework supports the assessment of long-term policy effects on economic performance, environmental sustainability, and market stability.

\begin{figure}[ht]
	\centering
	\includegraphics[scale=0.6]{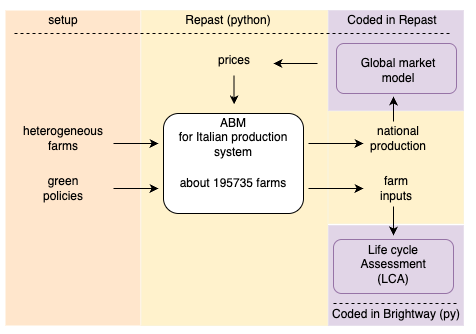}
	\caption{Visual representation of the software bundle.}
	\label{fig:arch}
\end{figure}

Figure \ref{fig:arch} gives a visual representation of the Integrated Model.

The software implementing the three modules is developed using the Repast Suite(\url{https://repast.github.io}) for the ABM and the GEM, and Brightway (\url{https://docs.brightway.dev/en/latest/}) for the LCA module.

\subsection{ABM setup: Identifying Heterogeneous Farms Using FADN Data}

The Agent-Based Model (ABM) relies on a detailed representation of the diversity within Italian wheat farms to simulate realistic production and policy responses. To construct this heterogeneity, we use microdata from the RICA (Réseau d’Information Comptable Agricole) database,
collected by the Italian Council for Agricultural Research and Agricultural Economy Analysis (CREA).
This dataset provides hundreds of variables, including farm balance sheets, resource use, and structural characteristics.

Filtering for durum wheat producers yields approximately 1,800 farms per year in the RICA sample.

To extrapolate this information to the national level, where the 2020 Agricultural Census reports 195,735 durum wheat farms (\cite{istat2022} p. 12), we build a synthetic population that reflects the sample's diversity while matching the real population size (\cite{rofasss25}).

We identify the main types of production systems through statistical clustering, capturing differences across four dimensions: 1) cultivated area (ha), 2) machinery use (hours of tractor work per hectare), 3) fertilizer application (kg of nitrogen, phosphorus, and potassium per hectare), and 4) pesticide usage (toxicity levels and amounts of herbicides, insecticides, and fungicides). Before clustering, we remove extreme outliers.

Each cluster represents a distinct production system. We assign weights to these clusters based on their representation in the RICA sample and scale them to generate a full synthetic population of 195,735 farms. For each artificial farm, input variables are sampled from statistical distributions estimated for its cluster, preserving realistic variability.

This approach reduces distortions from directly scaling the raw sample and ensures the ABM captures both representative farm types and the variability within them. The synthetic population forms the basis for simulating production decisions, yield responses, and adoption of policy measures at the national scale

\subsection{Farms' decision process and the complexity of the system}

In the ABM, each artificial farm is endowed with a decision-making process to determine its target wheat yield per hectare. The process follows basic microeconomic principles, where the farm maximizes profit by balancing the marginal revenue from higher yields against the marginal cost of additional inputs. Starting from a potential yield (the maximum attainable under optimal conditions), the model accounts for stress factors such as nutrient limitations, weeds, and pests. Fighting these stress factors raises yields but also increases costs, so the optimal target yield is identified by equating marginal cost and marginal revenue. Once the target yield is set, the required inputs are computed accordingly.

For a single stress factor, the target yield ($\hat{y}^*$) and input level ($\hat{x}^*$) 
can be derived analytically using the relationship between yield, input costs, and prices:
$$\hat{y}^*=\bar{y}-p_x/(p_w\lambda) \qquad \textnormal{and} \qquad
\hat{x}^*=-\frac{1}{\lambda}\ln\left(p_x/(p_w\lambda s\bar{y})\right)$$
where $\bar{y}$ is potential yield, $s$ is the share of $\bar{y}$ lost due to the stress factor, $\lambda$ regulates the gain in yield as fighting the stress, $x_i$ is the strength of the measure taken to contrast the stress, $p_w$ is the wheat price and $p_x$ is the price of $x$. For example, if the considered stress factor is plant nutrition, then $x$ is the quantity of fertilizer applied, and $p_x$ is the unit price of the fertilizer.
When multiple stress factors are present, solutions are computed numerically, producing the target yield and a separate input allocation for each factor.

Large farms are assumed to adopt this economically rational decision-making, while smaller or less sophisticated farms are assumed to use simpler, behavior-based principles. In addition, we should recognize that these decisions are taken before sowing, but actual yields often diverge from targets due to idiosyncratic shocks during the growing season. Therefore, the attained yield usually diverges from the target yield.
This enables the calculation of actual yields and, consequently, yield gaps—defined as the difference between potential and realized yields \cite{FISCHER20159,yielgapclimatalk,VANITTERSUM20134}.

The agent-based framework will be used in the future to incorporates interactions among farms, including information exchange on innovations and participation in voluntary policy schemes or local advisory programs. These elements enhance the realism of the system, allowing it to better reflect the complexity of actual farming dynamics.

\section{The global economic model (GEM)}

A central element in the farm decision process is the wheat price ($p_w$), which is typically determined in international markets. The GEM is used to compute these prices endogenously by simulating the dynamics of global wheat supply and demand

The GEM is based on the Commodity Markets Simulator (CMS) framework \cite{giulioni_novel_2019} \cite{giulioni18}, calibrated using FAOSTAT data. It outputs wheat prices across twelve international markets	and traded wheat quantities in twenty-four world regions.

Agents in the GEM are either producers or buyers, representative of large geographic areas (normally aggregation of several countries), who interact to negotiate wheat.
More specifically, we use FAOSTAT  sub-continental geographic areas. However, we split areas that include countries playing a relevant role in the world wheat production/consumption system and consider those countries as additional agents. Since the focus of this work is on the Italian wheat production system, we also allow Italy to be an additional agent.
The global wheat market is driven by a few key producers: China, India, Russia, and the United States, while Italy ranks 21st. Traditional major exporters include the United States, Canada, and Australia, with Russia and Ukraine emerging as dominant players in recent decades. Since 2002, Italy has had a structural deficit of about 2.5 million tonnes annually, driven by a large domestic pasta and wheat-processing industry.

In the GEM, buyers and sellers exchange offers in market sessions until agreements are reached. Each producer participates in at least one session, and the equilibrium price and traded quantity are derived from aggregated supply and demand curves. Once deals are settled, goods move directly from sellers to buyers, with transportation costs borne by the buyers.

By feeding international price signals back to the ABM and receiving updated national production levels, the GEM closes the feedback loop between global markets and farm-level decisions, enabling consistent seasonal simulations

\section{LCA computation}

The environmental impacts of wheat production are quantified using a Life Cycle Assessment (LCA) approach, in accordance with ISO 14040 standards. The analysis focuses on inventory compilation and impact assessment, quantifying the material and energy flows of agricultural operations and translating them into environmental indicators.

The LCA is implemented using Brightway2, an open-source Python platform for life cycle analysis. Data from the RICA dataset, which include fertilizer and pesticide use, machinery operations, and fossil fuel consumption, are processed with the ReCiPe 2016 impact assessment framework  \cite{Huijbregts2016,Huijbregts2017}. This methodology evaluates multiple impact categories such as greenhouse gas emissions, soil acidification, ecosystem degradation, and resource depletion.

\begin{figure}[ht]
	\centering
	\includegraphics[scale=0.5]{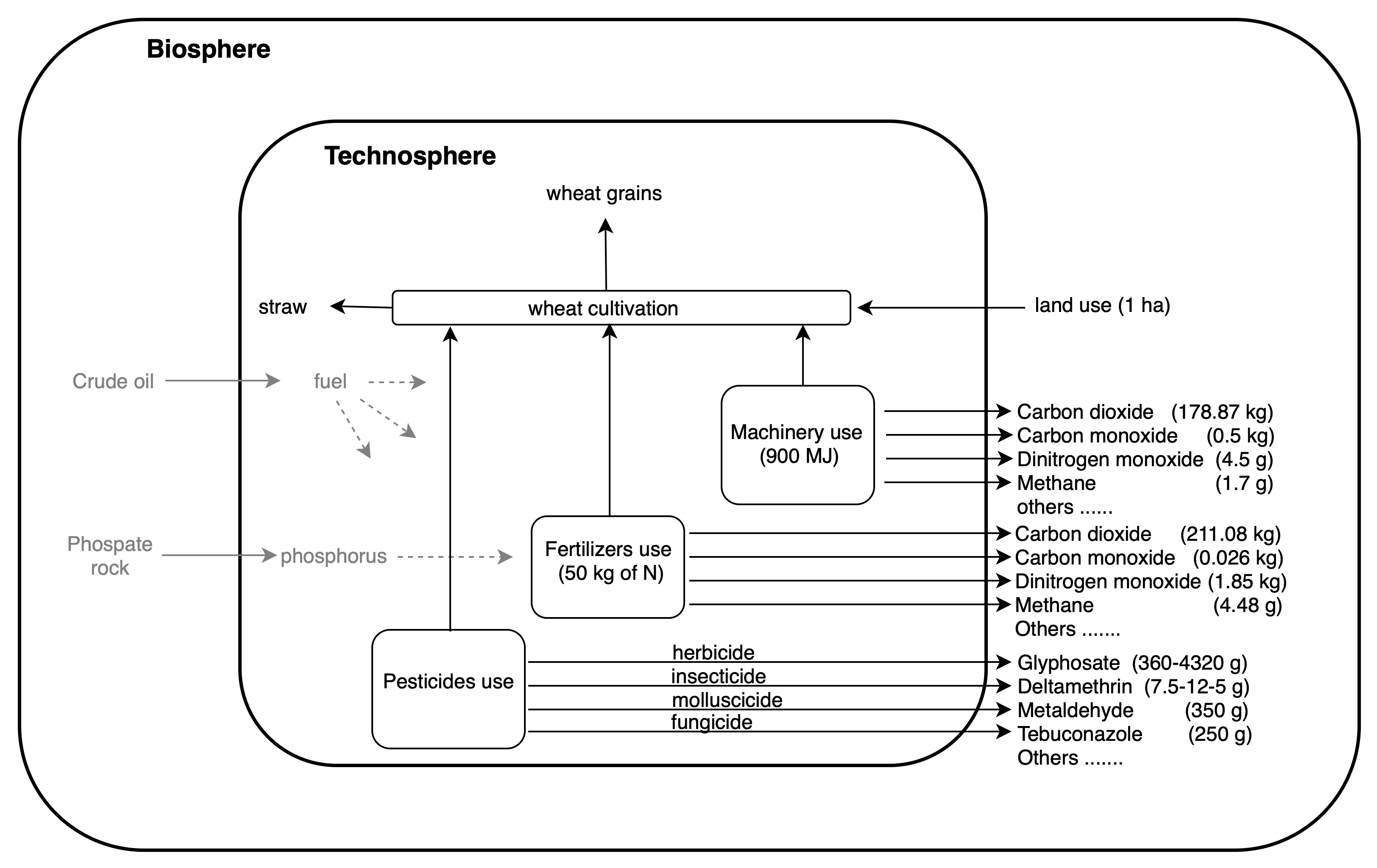}
	\caption{Visual representation of Life Cycle Inventory phase of the wheat production process.}
	\label{fig:lca}
\end{figure}

The assessment emphasizes endpoint indicators, which summarize environmental impacts into decision-relevant metrics like Disability-Adjusted Life Years (DALYs), species loss, and economic costs. While endpoint indicators introduce greater uncertainty due to value choices and modeling assumptions, they provide a clear synthesis for policymakers and non-experts, supporting trade-off evaluations and strategic planning.

The LCA is applied to every farm in the ABM after each simulation cycle, producing farm-level and aggregated national indicators (see Figure \ref{fig:lca} for a visual representation of the inventory phase performed at the farm level). These outputs allow the integrated framework to link economic and environmental performance, helping evaluate the sustainability of policy scenarios for Italian wheat production.

\section{Discussion}
This work presents a software framework integrating an Agent-Based Model (ABM), a Global Economic Model (GEM), and a Life Cycle Assessment (LCA) module to evaluate how policy changes and shocks affect the Italian wheat production system. The system focuses on assessing environmental impacts while capturing dynamic interactions between local farm-level decisions and international markets.

A strength of this approach is its capacity to capture heterogeneity at the farm level—simulating nearly 196000 farms—while incorporating global price mechanisms. By iterating across multiple seasons, the framework reveals how green policy incentives propagate through both economic and ecological systems.

\begin{figure}[ht]
	\centering
	\includegraphics[scale=0.13]{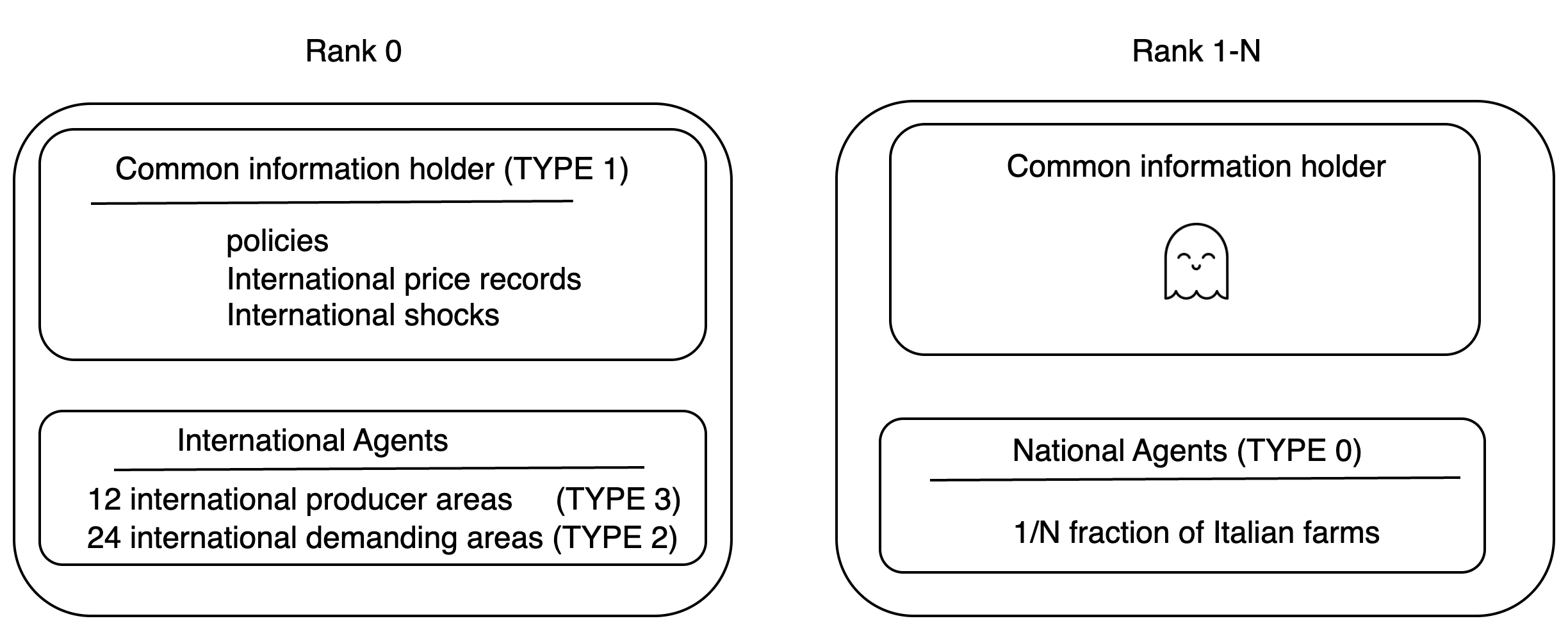}
	\caption{Visual representation of the agents in our parallel implementation. The common information holder is available in ranks 1-N as a copy of the agent in rank 0, i.e., a ghost in repast4py terminology}
	\label{fig:agents}
\end{figure}

A central technical challenge of the project is the computational scalability, as the farms decision process becomes more complex, simulating a system with large number of agents could require high-performance computing (HPC). To address this, we decided to implement the Italian wheat model in the Repast for Python (\url{https://repast.github.io}), leveraging mpi4py to parallelize simulations. Accordingly, international market computations run on the master processor (rank 0), while the Italian ABM is distributed across the other ranks. This approach significantly reduces runtime and enables large-scale scenario analysis.
Figure \ref{fig:agents} gives a visual representation of the agents populating the framework.

The integrated framework is designed to evaluate policy impacts under various conditions, including the rollout of the European Common Agricultural Policy (CAP), substitution of traditional policies with green incentives, and the occurrence of unexpected shocks. By combining economic and environmental assessments in a scalable, parallelized system, the model offers policymakers a tool to explore trade-offs and synergies between productivity, market stability, and sustainability in the Italian wheat sector

However, there are limitations worth noting. First, the model assumes rational optimization in large farms and rule-based approximations in others, potentially oversimplifying behavioral responses. Second, LCA results—especially when focusing on endpoint indicators like DALYs or species loss—carry uncertainty due to assumptions in value choices and ReCiPe parameterization. Third, calibration hinges on RICA/FADN and FAOSTAT data, which may not fully capture future trends or changes in agricultural practices.

Despite these caveats, the integrated tool offers policymakers a synthetic yet systematic way to evaluate trade-offs between production, price stability, and environmental outcomes in a global-local context.


\section*{Acknowledgments} This research was conducted as part of the project "ECOWHEATALY: Evaluation of policies for enhancing sustainable wheat production in Italy" (202288L9YN) funded by the European Union-Next Generation EU under the call issued by the Minister of University and Research for the funding of research projects of relevant national interest (PRIN) - "PRIN 2022".

%
%
%
 \bibliographystyle{splncs04}
 \bibliography{references}
\end{document}